# RDCL 3D, a Model Agnostic Web Framework for the Design and Composition of NFV Services

Stefano Salsano[1,2], Francesco Lombardo[1], Claudio Pisa[1], Pierluigi Greto[1], Nicola Blefari-Melazzi[1,2]

(1) CNIT, Italy - (2) University of Rome Tor Vergata, Italy

*Abstract*—We present RDCL 3D, a "model agnostic" web framework for the design and composition of NFV services and components. The framework allows editing and validating the descriptors of services and components both textually and graphically and supports the interaction with external orchestrators or with deployment and execution environments. RDCL 3D is open source and designed with a modular approach, allowing developers to "plug in" the support for new models. We describe several advances with respect to the NFV state of the art, which have been implemented with RDCL 3D. We have integrated in the platform the latest ETSI NFV ISG model specifications for which no parsers/validators were available. We have also included in the platform the support for OASIS TOSCA models, reusing existing parsers. Then we have considered the modelling of components in a modular software router (Click), which goes beyond the traditional scope of NFV. We have further developed this approach by combining traditional NFV components (Virtual Network Functions) and Click elements in a single model. Finally, we have considered the support of this solution using the Unikernels virtualization technology.

## I. INTRODUCTION

Traditional networks rely on physical appliances such as routers, switches, firewalls and load balancers, which are specialized for specific functions and are implemented on dedicated hardware. The Network Function Virtualization (NFV) paradigm aims at porting these appliances to software, creating Virtualized Network Functions (VNFs) to be executed on general-purpose devices. Then Network Services (NSs) can be realized by chaining and composing VNFs. This allows for a greater flexibility, efficiency in terms of resource utilization, modularity, and extensive automation.

In the NFV paradigm, the NFV Orchestrator (NFVO) plays a critical role. The NFVO instantiates Network Services combining the VNFs and interacting with the Virtual Infrastructure Managers (VIMs) that control the NFV Infrastructure (NFVI). To this aim, the Orchestrator needs as input a representation of services and components (VNFs) expressed with proper description languages / data models. The definition of such languages and data models is an ongoing process, to which several actors are contributing: Standard Defining Organizations (e.g. ETSI [1], OASIS [2]), open communities (e.g. OPNFV [4]) and other organizations that are developing NFV Orchestrators like OSM [5], OPEN-O [6] and Open Baton [7]. Each of the above-mentioned actors is currently working on a different language / data model for NFV.

This work was performed in the context of the project Superfluidity, which received funding from the European Union's Horizon 2020 research and innovation programme under grant agreement No. 671566.

The Superfluidity research project [8][9] fully embodies the NFV concepts and advances them considering the support of heterogeneous infrastructures and execution platforms and the decomposition of services into smaller reusable components.

In this context, we propose the RDCL 3D framework, a tool for the design, composition and deployment of virtualized network services. Compared to related work, RDCL 3D targets a broad and evolving set of network service description languages, emphasizes nested composition and is conceived to be flexible and extensible.

In section II, we introduce the *NFV MANO Reference Architecture*, while in section III we generalize its concepts, introducing the concept of *Reusable Functional Blocks* (RFBs) and providing the notions of *RFB Execution Environment* (REE) and *RFB Composition and Definition Language* (RCDL). Then in section IV we describe the functionality of the RDCL 3D framework and the different options for integrating it into the NFV MANO Architecture. Section V discusses a set of applications implemented using the platform, showing how it can support innovative NFV solutions going beyond the state of the art of NFV models and orchestration tools. Section VI provides details of the software architecture of the framework and illustrates how new models can be "plugged in". Section VII reports the related work, while section VIII contains the conclusions and a recap of the main contributions of the paper.

## II. NFV MANO REFERENCE ARCHITECTURE

In the last three years, the ETSI NFV Industry Specification Group (ISG) [1] has released a number of documents describing the *management and orchestration* (MANO) framework of *Virtual Network Functions* (VNFs). Fig. 1 depicts an overview of the architecture proposed by ETSI [10]. The VNFs rely on the resources provided by the *Network Functions Virtualisation Infrastructure* (NFVI) layer, which includes compute, networking and storage hardware resources. These resources are managed by one or more *Virtualised Infrastructure Managers* (VIMs), which expose northbound interfaces to the *VNF Managers* (VNFM) and to the *Network Functions Virtualisation Orchestrator* (NFVO).

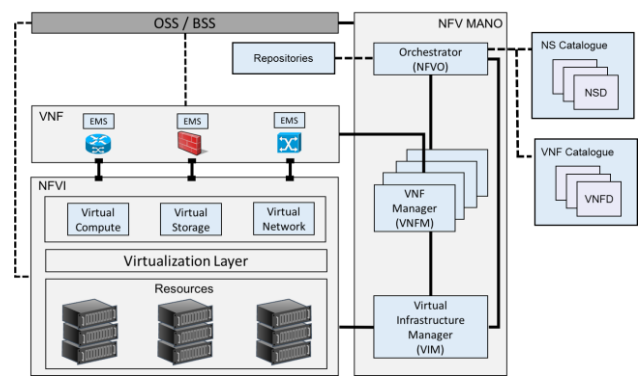

Fig. 1. ETSI NFV MANO Reference Architecture diagram

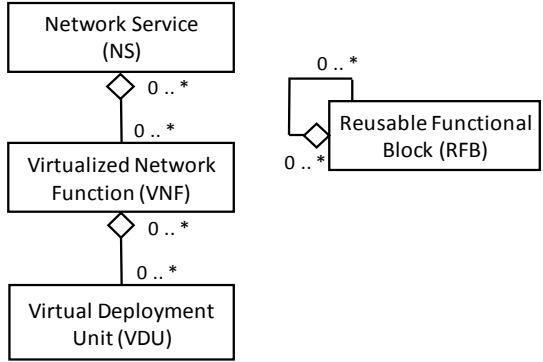

Fig. 2. ETSI MANO (left) vs. Superfluidity proposed model (right)

The VNFM performs the lifecycle management of the VNFs. The associated *VNF Catalogue* stores the *VNF Descriptors* (VNFDs), which describe the structural properties of the VNFs (e.g. number of ports, internal decomposition in components) and their deployment and operational behaviour. A VNF is decomposed in one or more *Virtual Deployment Units* (VDUs) which can be mapped to Virtual Machines (VMs) or containers to be deployed and executed over the NFV Infrastructure. The NFVO is responsible for the overall orchestration and lifecycle management of the *Network Services* (NSs), which combine VNFs according to an interconnection topology. A Network Service is represented by an *NS Descriptor* (NSD), which captures the relationship between VNFs. The NSDs are stored in an *NS Catalogue* and are used by the NFVO during the deployment and operational management of the services. The NFVO is also responsible for the "on-boarding" and validation of the descriptors of VNFs and NSs.

To sum up, the network operator is able to build services (NSs) combining Virtual Network Functions (VNFs) using an orchestrator (NFVO). The NFVO is able to process the descriptors of services (NSDs) and of VNF (VNFDs) and to interact with the manager (VIM) of the virtualized infrastructure (NFVI) in order to deploy the components that implement the service and interconnect them appropriately. In particular, the descriptors of the virtual network functions include the Virtual Deployment Units (VDUs), which are mapped into Virtual Machines or Containers that can be deployed over the deployment and execution infrastructure (NFVI).

### III. THE SUPERFLUIDITY ABSTRACTIONS: RFB, REE AND RDCL

The H2020 Superfluidity 5G-PPP project [8] [9] proposes a generalization and an extension of the architectural models described in section II. This section discusses the main concepts introduced by the project, showing the relationships with the ETSI models. A *Reusable Functional Block* (RFB) is a generalization and an abstraction of VDUs, VNFs and NSs. In the ETSI NFV MANO reference architecture, there is a fixed hierarchy among these three elements. The VDU is a component that can be deployed as a VM or Container over an NFV Infrastructure. VDUs are combined to form VNFs. VNFs are combined to form Network Services (Fig. 2, left). The ETSI reference architecture assumes that the execution environment is a virtualized infrastructure that can host VMs or containers and support their interconnection. On the other hand, the proposed RFB concept is more general: an RFB can be described in terms of other "nested" RFBs and this decomposition can be iterated arbitrarily, down to much "smaller" components (Fig. 2, right).

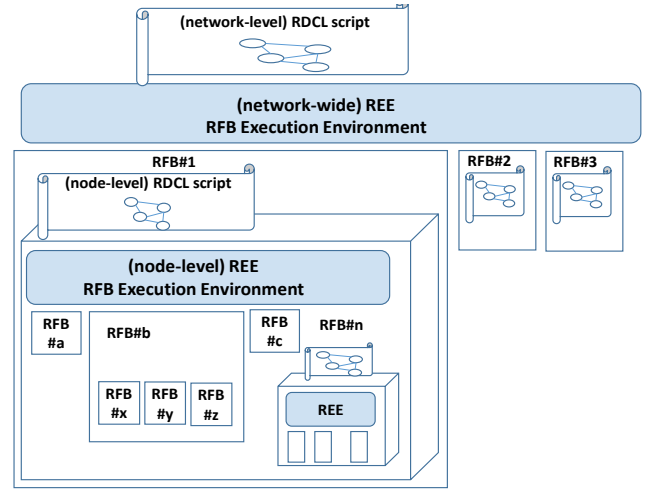

Fig. 3. RFB, RDCL and REE abstractions proposed by Superfluidity

In the Superfluidity model, different execution environments can support the deployment and the composition of RFBs, depending on the context. Hence, the key concept of *RFB Execution Environment* (REE) is introduced. In order to consider a concrete example, the virtualized infrastructure (NFVI) introduced in the ETSI architecture is mapped to a REE in the proposed modelling abstraction (the "network-wide" REE in Fig. 3). The VNFs are mapped into RFBs. Let us assume that a VDU in a VNF runs a software modular router like Click [11]. The Click modular router allows to compose the node functionality by combining Click *elements* in a *configuration*. In our model, the Click router is an RFB executed in the NFVI REE. This RFB is further decomposed in "smaller" nested RFBs (the Click elements). Hence, the Click modular router platform is the RFB execution environment (REE) for the nested RFBs. Looking at Fig. 3, RFB#2 and RFB#3 correspond to VNFs. RFB#1 corresponds to a Click platform, that is a "node level" REE. This REE hosts RFBs (#a, #b, #c…) which in turn can be composed by other RFBs (like RFB#b) executed in the same REE. This process can be further iterated, as shown for RFB#n, which is mapped in a third REE. For example this last REE could represent a hardware accelerated packet processor or a software radio board.

An RFB Execution Environment does not only represent the abstraction of the actual execution environment but includes also the tools that are used to deploy and run components/ services. For example, considering the ETSI MANO reference architecture, the REE corresponds to the virtualized infrastructure NFVI as actual execution environment and the REE tool is the orchestrator (NFVO). Considering the Click modular router as REE, the Click platform is the actual execution environment, while the software that parses the configuration files and produces a Click executable is the REE tool.

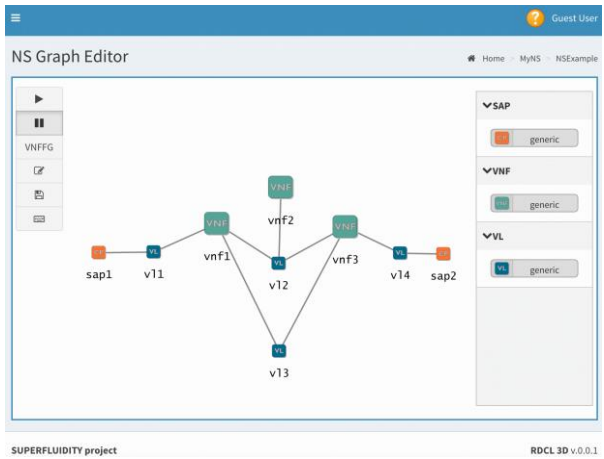

Fig. 4. Screenshot of RDCL 3D (design of a Network Service)

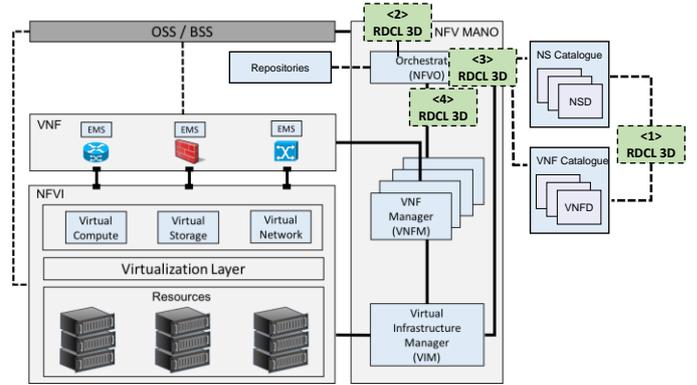

Fig. 5. Positioning RDCL 3D in the ETSI MANO architecture

The REE tools rely on languages to describe the components (RFBs), their requirements on the execution platform and their mutual interactions and composition to build other components or final services. We refer to these languages in general as *RFB Description and Composition Languages* (RDCLs). The *RDCL Scripts* are the instances of the description of services and/or components within a given REE. An RDCL Script describes the structure and behaviour of RFBs and how to compose them to form aggregated RFBs.

Considering the ETSI MANO architecture, the RDCL corresponds to the VNFD and NSD descriptors. We note that besides the VNFD and NSD standard descriptors proposed in [12][13] there are other RDCL languages in the same class, which are aligned in their goals and share the architectural approach but are not compatible in their syntax. These "sibling" languages are proposed by other SDOs or by organizations that develop NFV Orchestrators [2][5][6][7].

Considering the Click modular router platform, we move to a different class of RDCLs. The Click configuration language allows describing the arbitrary composition of packet processing modules (Click *elements*) provided in a catalogue (e.g. a library). It is possible to add new elements to the catalogue composing existing ones or to develop new "basic" elements (using the C++ programming language). The Click platform is further described in Section V.C.

The vision of the Superfluidity project is to orchestrate functions dynamically over and across heterogeneous environments, hence the need to understand and operate with different RDCL languages. At least in the short-medium term, we can envisage that the different RDCLs will coexist and that "meta-orchestrators" will coordinate the mapping of the configurations into resources over the different RFB Execution Environments involved. As we will show in section V.D, the existing RDCL languages may need to be extended to support the interaction with different REEs.

## IV. THE RDCL 3D FRAMEWORK AND USAGE SCENARIOS

In order to realize the architectural vision described in the previous section, we designed and implemented the RDCL 3D framework. RDCL 3D stands for *Reusable Functional Blocks Description and Composition Language Design, Deploy and Direct*. It is a novel "model agnostic" Web framework for the design of descriptors (i.e. RDCL Scripts, introduced above) which can be used for the deployment and operational management of NFV services and components (e.g. VNFs).

RDCL 3D offers a Web GUI that allows visualizing and editing the descriptors of components and network services both textually and graphically. A virtualised network service designer can create new descriptors or upload existing ones for visualization, editing, conversion or validation. The created descriptors can be stored online, shared with other users, or downloaded in textual format for use in other tools. Moreover, RDCL 3D includes the concept of *Agents* that can interact with external entities (e.g. Orchestrators or Infrastructure Managers) in order to deploy the service corresponding to the descriptor files. Fig. 4 shows a screenshot of RDCL 3D: the user can perform intuitive drag-and-drop operations on the graph representations of RDCLs; in this case a Network Service view showing the interconnection of VNFs through *Virtual Links*.

We have released the source code of RDCL 3D [14] under the Apache 2.0 Open Source license to facilitate the uptake by the research and industrial communities. An alpha release of the system is on line at [15]. On the online portal, it is possible to login as a guest to explore the RDCL 3D capabilities. It is also possible to request an account that allows saving and retrieving projects and related descriptor files across different sessions.

RDCL 3D is designed for extensibility in order to support new models or to combine existing models and to interact with different orchestrators or deployment platforms. Hence, we say that RDCL 3D is "model agnostic". The specific models and platforms that are currently supported are detailed in section V.

With reference to the ETSI MANO architecture, the RDCL 3D framework can be adapted to play different roles, as shown in Fig. 5. Hereafter, we describe three usage scenarios, which we have already implemented and a fourth case that can be potentially covered by the platform.

RDCL 3D can be used as a standalone tool to edit the NS and VNF descriptors (see <1> in Fig. 5). Under this approach, the produced descriptor files can be manually retrieved and provided to an Orchestrator (NFVO). This approach can be tested on the public on line demo for different types of models (e.g. the ETSI and TOSCA models).

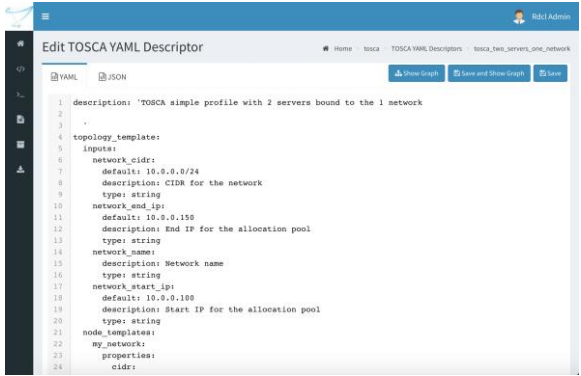

Fig. 6 Screenshot of the textual editing of a TOSCA simple profile descriptor in RDCL 3D.

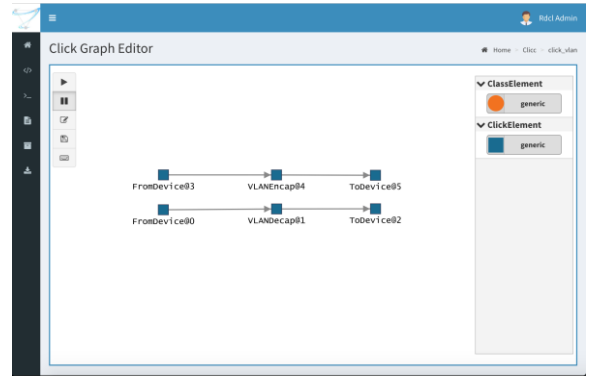

Fig. 7 Screenshot of the graphical editing of a Click modular router through the composition of its elements in RDCL 3D

RDCL 3D can support the direct interaction with programmatic APIs of external Orchestrators (<2> in Fig. 5), by including the logic for such interaction in an *Agent* module. In this case, the Agent module receives instructions from the RDCL 3D web GUI, hands the descriptor files to the external Orchestrator, and provides feedback to the user on the GUI. A specific use case on which we are working is the interaction with the ManageIQ [18] orchestrator platform. Note that this second scenario can be extended when there is the need to combine different orchestration platforms with different description languages, so that RDCL 3D can become a meta-orchestrator.

The third usage scenario that we have considered is to use the platform to build Orchestrator prototypes, so that the RDCL 3D plays the role of the NFVO (<3> in Fig. 5). This is especially useful when one needs to explore a new functionality and it is easier to have a small stand-alone proof-of-concept implementation rather than integrating the new functionality into a fully-fledged NFVO. In particular, we have implemented the interaction with the OpenVIM [17] Virtual Infrastructure Manager (released by ETSI in the context of the OSM project [5]) to deploy Unikernels [19], specialized minimalistic VMs.

A fourth potential usage for the proposed framework is to be integrated as a library in Orchestrators that do not yet support a GUI or to improve their existing GUI features (<4> in Fig. 5).

Finally, let us consider the objectives of the OPNFV *Models* project [16]: "convergence of information and/or data models related to NFV service/VNF management, as being defined in standards (SDOs) and as developed in open source projects". In this respect, a tool like RDCL 3D can help the research and standardization community to compare the different description languages and data models available today through practical implementations. In addition, we are working on tools to support the conversion of descriptors from one format to another.

## V. From traditional NFV Models To Superfluidity

In this section, we provide the details of the project types (and their corresponding models) that are currently supported in the RDCL 3D framework and we describe the Agents that we have developed to interact with infrastructure managers. We first describe two sibling models for the description of traditional NFV scenarios based on VMs/containers (referred to as *ETSI* and *TOSCA* models), and then we consider the support of a model for granular decomposition of a software router in packet processing elements (*Click* model). Finally, we propose a model that combines the *ETSI* traditional NFV model with the *Click* one, by means of a minor extension to the former one. In order to deploy the proposed solution using the Unikernels virtualization technology [19], we have extended the OpenVIM [17] Virtual Infrastructure Manager and developed an RDCL 3D Agent to interact with the extended OpenVIM.

### A. ETSI NFV ISG model

We have implemented from scratch the parsing, visualization and editing of the ETSI NFV ISG model descriptors, compliant with the latest specification released in October 2016 [12][13].

RDCL 3D can fully handle the visual and textual (in JSON and YAML formats) editing of VNFDs and NSDs, including *VNF Forwarding Graphs* (VNFFG), which capture the forwarding path of traffic flows in a Network Service.

### B. TOSCA simple model for NFV

The *Topology and Orchestration Specification for Cloud Applications* (TOSCA) [20] is an OASIS standard for the description of service components, their topology and their management processes. The "TOSCA Simple Profile in YAML" [3] provides the language specifications, while the TOSCA NFV profile [2] specializes this language for the NFV scenario. In RDCL 3D, both TOSCA profiles are supported by the integration of the OpenStack TOSCA parser [21] and Heat-Translator [22] tools. RDCL 3D can handle the visual and textual editing of TOSCA profiles and the generation of Templates for OpenStack Heat, i.e. the main project in the OpenStack Orchestration program. Fig. 6 shows a screenshot of the textual editing of a TOSCA simple profile descriptor.

### C. Click modular router model

The Click Modular Router software architecture [11] allows building flexible and configurable routers on Linux. A developer can define a Click Router instance as a composition of packet processing modules called Click Elements. These elements have directional input and output ports, which can be linked to other elements forming a connection graph. Examples of functions performed by Click elements are packet classification, queueing and scheduling. Click employs its own *Configuration* language, in which new Click Elements can be defined in a *Click Configuration* file as the composition of other Click Elements. The language can also be extended by developing new elementary Click elements in the C++ language. Click router instances are normally executed as Linux applications. However, Click has been successfully integrated in a Unikernel [23] environment called ClickOS [24], so that Click instances can be deployed as tiny VMs in an NFVI.

RDCL 3D supports the graphical visualization of Click configurations and their textual and graphical editing. Fig. 7 shows a screenshot of the graphical editing of a Click modular router.

*D. Superfluidity: combining ETSI NFV ISG and Click models*

Section III has provided a description of some of the concepts introduced by Superfluidity and especially the iterative composition of heterogeneous VNFs generalized by the idea of Reusable Functional Blocks. In order to provide a concrete implementation of this approach, we extended the VDU information element contained in the VNF Descriptors to reference Click router configurations (and potentially other types of Execution Environments that can be instantiated within a VDU and described by means of some descriptor). In particular, we have introduced two new attributes to the VDU information element defined in [12]: namely *vduNestedDescType* and *vduNestedDesc* (see Table 1). The *vduNestedDescType* attribute defines the type of RFB Execution Environment that is running in the VDU (in our case *Click*). The *vduNestedDesc* is an identifier. It provides a reference to the actual descriptor that is deployed in the REE running in the VDU (in our case a Click *configuration file*). The proposed *vduNestedDesc* uses the same approach of the *swImage* attribute in [12], which provides a reference to the actual software image that is deployed in a VDU.

We have also introduced a new attribute, namely *internalIfRef* (see Table 2), to the *VduCpd* information element. The *VduCpd* information element is also defined in [12] and is referenced by the VDU information element through the *intCpd* attribute. We add the attribute *internalIfRef* to the VduCpd element to create a correspondence between a VduCpd element and the network interface of a multi-interface VDU. For example, ClickOS instances internally name their interfaces as numbers starting at "0". A ClickOS-based firewall with two interfaces would thus have an interface named "0" and an interface named "1". The firewall could expect (in its Click configuration file) traffic from an external network A on port "0" and traffic from an internal network B on port "1". The VDU corresponding to this ClickOS-based firewall would thus have two internal VDU connection points, one leading (through other connection points and virtual links) to the network A and one leading to network B. In this case the *VduCpd* element that would lead to network A would have its *internalIfRef* attribute set to "0", while the *VduCpd* element that would lead to network B would have its *internalIfRef* attribute set to "1".

In principle, the proposed new attributes give the possibility to include a formal description of the VDU internals according to different types of languages, in a flexible and extendible way. In particular, we have only focused on the use of Click configurations to expand further a VDU. This yields to a description language that encompasses the definition of high level NSs, the chaining of VNFs (taken from the ETSI specification) and the composition of low level Click Elements. We call this language the *Superfluidity-ETSI RDCL* and it represents a step in the direction of realizing the RFB abstractions and composition models described in Section III.

The RDCL 3D framework supports the Superfluidity-ETSI RDCL with the graphical visualization and nested editing of NSDs, VNFDs and Click configurations. The nested visualization is triggered on the web GUI by clicking on the nodes of the visualized graphs.

For what concerns the deployment of Superfluidity-ETSI network services, this can be performed by an *RDCL 3D Agent*. The RDCL 3D agents are implemented as backend component, they allow the deployment of network services through the interaction with NFVOs and/or VIMs and are further described in Section VI.

To demonstrate the feasibility of our approach, we have modified OpenVIM ([17]) in order to support the deployment of Unikernel/ClickOS-based VNFs. Moreover, we have developed an RDCL 3D agent, associated to the Superfluidity-ETSI model, which performs an online descriptor format translation and deployment leveraging the modified OpenVIM APIs.

## VI. RDCL 3D SOFTWARE ARCHITECTURE

RDCL 3D is a web application with a backend component and a frontend component (see Fig. 8). The backend component running on a web server is based on the Python Django framework. It includes the persistence layer (database). The frontend component running in the web browser is developed in Javascript and exploits D3.js, a library for interactive data visualization. The platform is designed with a modular approach both in the backend and in the frontend, so that it can be extended to support new project types with minor development effort. Each project type can be seen as a "plugin" for the RDCL 3D framework, composed by a backend plugin (in Python) and a frontend plugin (in Javascript). The backend plugin is split into a Project Handler plugin (mandatory) and a Deployment Handler plugin which is needed for the interaction with an Agent that talks to an external orchestrator or a VIM.

| Attribute | Qualif. | Cardin. | Content | Description |
|---|---|---|---|---|
| vduNestedDesc | O | 0..1 | Identifier (Reference to a VduNestedDesc) | This is a reference to the actual descriptor deployed in the VDU (e.g. a Click configuration). The reference can be relative to the root of the VNF Package or can be a URL. |
| vduNestedDescType | O | 0..1 | Enum | Identifies the type of descriptor file referred in the vduNestedDesc field (Click configuration, etc.) and consequently the type of the Execution Environment running in the VDU. This field must be present if vduNestedDesc is present. |

Table 1 New attributes for the ETSI MANO VDU Information element

| Attribute | Qualif. | Cardin. | Content | Description |
|---|---|---|---|---|
| internalIfRef | O | 0..1 | String | Identifies the network interface of the VDU which corresponds to the VduCpd. This attribute allows to bind the VduCpd to a specific network interface of a multi-interface VDU. |

Table 2 New attributes for the ETSI MANO VduCpd Information element

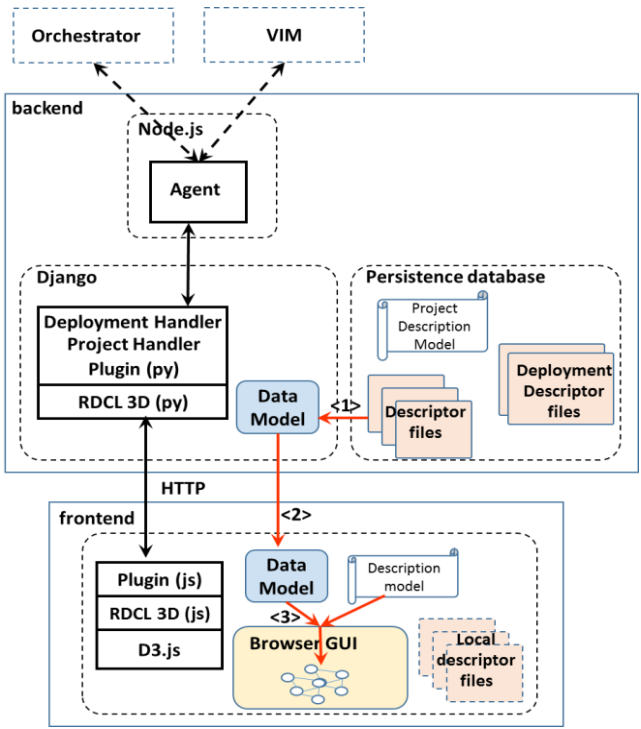

Fig. 8 RDCL 3D software architecture

Each user (e.g. a service designer or network administrator) can instantiate projects of the supported project types. The descriptor files for the projects are stored in the persistence layer in the backend. Predefined descriptor files are available for each project type (i.e. they represent examples of services or existing components that can be reused). A Data model for a project is created in the backend Project Handler plugin by parsing and validating the descriptor files (<1> in Fig. 8). This process is specific for the project type, therefore it is performed by the specific plugin for the project type. The instance of the Data model contains all the information of a project instance i.e. all the information contained in the project descriptor files. The Data model is sent to the frontend (<2> in Fig. 8), where it is processed and filtered to produce the different graphical views on the browser GUI (<3> in Fig. 8). The operations on the GUI (e.g. adding or removing nodes and links, editing of the local descriptor files) are reflected on the local version of the Data model and sent to the backend when it is needed to update the information stored in the persistence layer (e.g. the descriptor files). The backend can also optionally deploy the designed network services through an *RDCL 3D Agent*. An RDCL 3D Agent can deploy the designed network services by either interacting directly with the APIs of an Orchestrator or a VIM or by pushing the network service descriptors to a git repository to which the Orchestrator/VIM has access. Moreover, to allow the interaction with specific VIMs, the RDCL 3D Agent can also perform online translation between descriptor formats. The RDCL 3D Agent exposes REST APIs, which are invoked by the Deployment Handler plugin. Our RDCL 3D Agent implementations are based on node.js, but any technology which allows to expose HTTP REST APIs can be employed.

The operations performed by the frontend, like visualizing a view of the graph, adding/removing nodes and links, are dependent on the project type, so that they should be handled by the Javascript plugins. We are able to minimize the code that needs to be developed in a plugin to support a project type by introducing a *description model* for a project type. The *description model* includes the types of nodes and links that are supported, their relationships, the constraints in their composition, describes what are the different views of the projects and which nodes and links belongs to which view. The *description model* is expressed as a YAML file [25]. By parsing it, the Javascript frontend is able to perform most of the operations without the need of specific code for the project type. In order to handle the operations specific to the project type, the description model includes the possibility to associate operations to functions that are defined in the plugin. The definition of the structure of the description model and some examples are included in the *docs* folder in [14].

In order to simplify the integration of new project types, the framework includes a script that creates the skeletons of the Python and Javascript plugins (respectively for the backend and the frontend) and of the description model. Starting from these skeletons, a developer adding support for a new project would:

i) include the parser to translate the descriptor files into the Data model representation in the Python plugin (backend);

ii) customize the description model, capturing all the relevant properties of the project type and identifying the operations that need to be processes in a specific way for the project type by the Javascript plugin;

iii) develop the project type specific processing operation in the Javascript plugin (frontend);

iv) optionally, develop a deployment handler and an RDCL 3D Agent to allow the direct deployment of network services.

## VII. RELATED WORK

A comprehensive survey on NFV service orchestration aspects is provided in [26]. It would be unfeasible to provide here an exhaustive coverage of the large number of projects and activities ongoing in this area. For these reason we have selected and described below the projects that in our opinion are more closely related to our work.

ETSI OSM [5] is an NFVO, which includes a graphical composer for network services. Although conceptually aligned with the first release of the ETSI NFV ISG models, the descriptor syntax and some details are not compliant with the ETSI NFV ISG specifications. Compared to the ETSI OSM composer for network services, RDCL 3D targets a broader and evolving RDCL landscape which includes (but is not limited to) ETSI NFV ISG specifications as well as the combination of nested heterogeneous RDCLs.

Similarly, the Cloudify NFVO [27] is shipped with a graphical designer for TOSCA-based descriptors, but does not support other RDCLs and puts limited emphasis on composition of nested components.

OPEN-O [6] is a framework aligned with the ETSI NFV MANO architecture, which aims at orchestrating services on legacy networks as well as on SDN and NFV infrastructure. OPEN-O uses TOSCA templates for the description of network services and provides a graphical modelling designer. Its implementation relies on the Cloudify code base, thus from the RDCL point of view it is focused on the TOSCA language.

SONATA [28] is a H2020 5G-PPP EU-funded project. It aims at producing a development and orchestration framework for the NFV environment. The SONATA concept and architecture, like Superfluidity, also relies on the ETSI NFV ISG models, and it also encompasses the concept of recursiveness between NSs. However, SONATA does not address directly the description of the internal elements of, e.g., the VDUs. SONATA has released an open source Service Platform, which includes a Web GUI. This GUI targets the use cases of the project and it is not proposed as a general framework as the RDCL 3D GUI.

The management and orchestration framework Open Baton [7] is conceptually aligned with the first release of the ETSI NFV ISG models, while the syntax of the descriptors and the model details are not aligned. At the time of writing, Open Baton does not provide a graphical user interface for the design of network services.

## VIII. CONCLUSIONS

The first contribution of the paper is the description of the RDCL 3D, a web framework to visualize and edit services and components in NFV scenarios and to interact with NFV Orchestrators or Virtual Infrastructure Managers. RDCL 3D is not focused on a specific data model / description language. It is designed to facilitate the support and the integration of any model and language and the interaction with any Orchestrator/VIMs. In particular: i) RDCL 3D has a modular architecture in which a new project type can be added as a plugin; ii) a description model allows describing the structural properties of the project type, minimizing the need to develop code; iii) a script is used to generate the skeletons of the plugins and of the description model, to reduce the development effort. The RDCL 3D code has been released as open source.

Looking at the ETSI MANO reference model, we have shown how the RDCL 3D framework can be used in different scenarios to play different roles, for example: i) a standalone editor of NFV model descriptor files; ii) a GUI tool to interact with external Orchestrators; iii) a tool for fast prototyping of Orchestrator functionality interacting with VIMs.

We have provided a set of relevant contributions concerning the extension of traditional NFV models along with the prototype implementation of these extensions using the RDCL 3D framework. In particular, we have considered the support of heterogeneous execution infrastructures and the evolution towards more granular NFV models, allowing a higher abstraction in the description of components and their nested decomposition. We have provided a concrete example of how a traditional NFV model can be combined with a granular router decomposition model (Click) in a unified vision, with minor extensions to the traditional NFV model. We have also considered the deployment of this combined model onto an infrastructure supporting the Unikernel virtualization technology in parallel to the traditional VM based virtualization.

From the NFV design perspective, we have shown how the current models can be extended to support a more granular (and nested) decomposition of services into smaller components.

From the implementation experience, we claim that RDCL 3D is a highly versatile tool that facilitates the design and development of innovative NFV solutions. Thanks to its open source nature, the tool is available to the NFV research and development community.